\documentclass[12pt,a4]{article}
\usepackage{cite}
\usepackage{epsfig}
\begin{document}

\begin{center}
{\Large \bf Search for Exotics and Extra Dimensions at LEP} \\

\vspace{4mm}

Brigitte Vachon\\

\vspace{4mm}
University of Victoria \\
Department of Physics and Astronomy \\
P.O. Box 3055 \\
Victoria, B.C., Canada \\
V8W 3P6
\end{center}

\begin{abstract}
Results from various searches for new physical phenomena
performed by the four LEP experiments
are summarised.  Topics presented include the search for contact
interactions, a $\rm Z^\prime$ boson, leptoquarks, excited leptons,
technicolour and gravity in extra dimensions.

\end{abstract}

Many different theoretical ideas of physics beyond the Standard Model
attempt to address aspects of nature that remain unexplained. 
Searches for physical consequences of many
models of new physics are performed by the four LEP experiments:
ALEPH, DELPHI, L3 and OPAL.  This notes summarizes the most recent
preliminary results obtained in the search for four-fermion 
contact interactions, a
$\rm Z^\prime$ boson, leptoquarks, excited leptons,
technicolour and gravity in extra dimensions.

During the second phase of LEP running, 
each experiment recorded a total of approximately 600~$\rm pb^{\rm
-1}$ of data.  Table~\ref{table:lumi} shows a break down of the
approximate integrated luminosity recorded per experiment at each
centre-of-mass energy attained.  Not all results presented here include the
entire set of 
data; whenever possible the data set used in each searches is indicated.

The substantial amount of data recorded combined with the highest
centre-of-mass energies ever achieved in $\rm e^+e^-$ collisions
provide a unique environment to look for new phenomena beyond the
Standard Model.

\begin{table}[h]
\begin{center}
\begin{tabular}{|l|c|r|}\hline
year & $\sqrt{s}$ & \multicolumn{1}{|c|}{$\cal L$} \\ 
     & (GeV)      &  \multicolumn{1}{|c|}{($\rm pb^{-1}$)} \\ \hline\hline
1996 & 161-172    & $\sim 20$    \\ \hline
1997 & 183        & $\sim 40$    \\ \hline
1998 & 189        & $\sim 180$   \\ \hline
1999 & 192-202    & $\sim 230$   \\ \hline
2000 & 200-209    & $\sim 220$   \\ \hline
\end{tabular}
\end{center}
\caption{\label{table:lumi}Center-of-mass energies and corresponding
approximate integrated luminosities recorded per experiment during the second phase of LEP running.  }
\end{table}

\section{Four-Fermion Contact Interactions}
The mathematical formulation of four-fermion contact interactions is a 
general framework for
describing possible deviations from the Standard Model interactions.
At LEP, the experiments search for contact interactions of the 
form $\rm e^+e^-f\bar{\rm f}$.
The general gauge invariant and chiral/flavour conserving Lagrangian
used to describe possible contact interactions is given by~\cite{bib:ci_models}

\[{\cal L}_{\rm contact} = \frac{g^2}{(1+\delta)\Lambda^2}
\sum_{ij=\rm L,R}\eta_{ij}(\bar{ e}_i\gamma^\mu e_i )
(\bar{f}_j\gamma_\mu f_j ) \]

\noindent
where $g^2$ is usually set arbitrarily to $4\pi$ by convention,
$\Lambda$ is the
energy scale of these new interactions, $\delta = 1$ only for eeee
interactions and is zero otherwise, and $e_i,f_i$ are left or right-handed
spinors.  The parameters $\eta_{ij}$, which have values
generally between -1 and 1, describe the chiral 
structure of the interactions . The different 
models~\cite{bib:ci_models} considered and their corresponding values
of $\eta_{ij}$ are presented in table~\ref{table:ci_models}. 

\begin{table}
\begin{center}
\begin{tabular}{|c|r r r r|} \hline
Model & $\eta_{\rm LL}$ & $\eta_{\rm RR}$ & $\eta_{\rm LR}$ &
$\eta_{\rm RL}$ \vspace{-0.2cm} \\ 
name & & & & \\ \hline
$\rm LL^\pm$ & $\pm 1$ & 0 & 0 & 0 \\ 
$\rm RR^\pm$ & 0 & $\pm 1$ & 0 & 0 \\
$\rm LR^\pm$ & 0 & 0 & $\pm 1$ & 0 \\
$\rm RL^\pm$ & 0 & 0 & 0 & $\pm 1$ \\
$\rm AA^\pm$ & $\pm 1$ & $\pm 1$ & $\mp 1$ & $\mp 1$ \\
$\rm VV^\pm$ & $\pm 1$ & $\pm 1$ & $\pm 1$ & $\pm 1$ \\
$\rm V0^\pm$ & $\pm 1$ & $\pm 1$ & 0 & 0 \\
$\rm A0^\pm$ & 0 & 0 & $\pm 1$ & $\pm 1$ \\ \hline
\end{tabular}
\end{center}
\caption{\label{table:ci_models}Types of four-fermion contact
interactions considered and their corresponding values of $\eta_{ij}$.
For each model, both constructive (+) and destructive (-) interference
between the Standard Model processes and the contact interactions are
considered. }
\end{table}

\begin{figure}
\begin{center}
\raisebox{-1.ex}{\epsfig{file=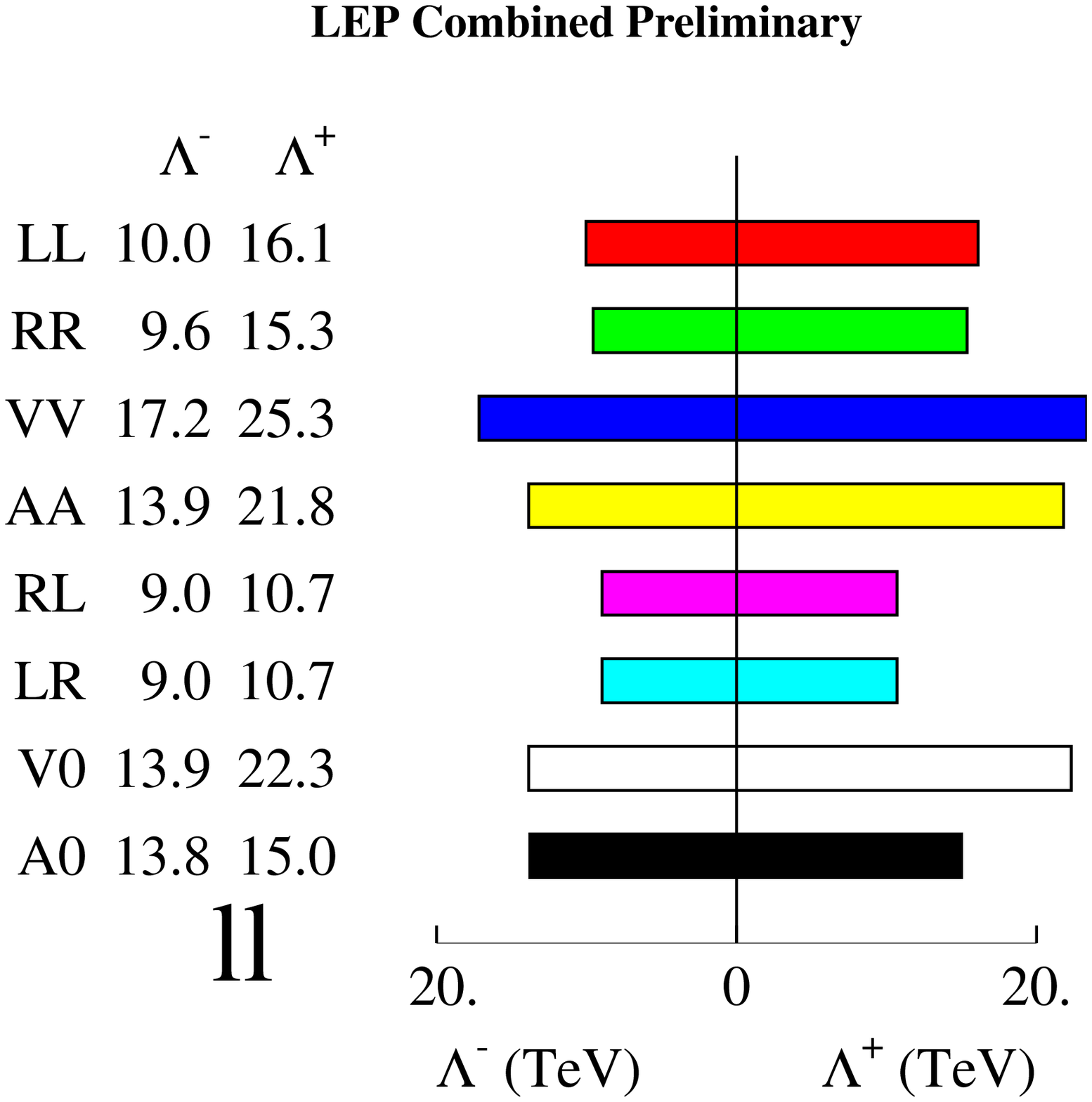,height=6.5cm}} 
\mbox{\epsfig{file=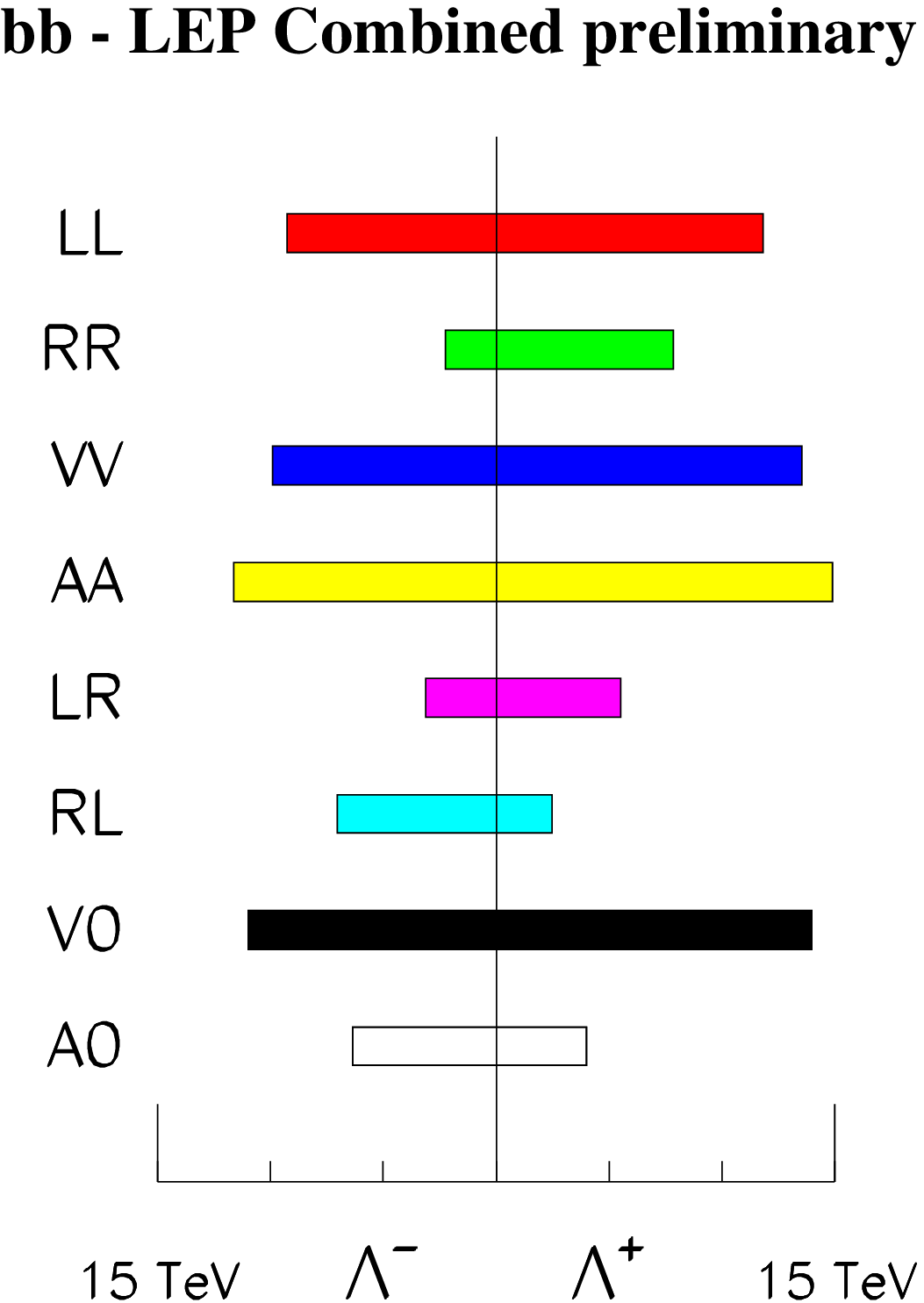,height=6.cm}}
\mbox{\epsfig{file=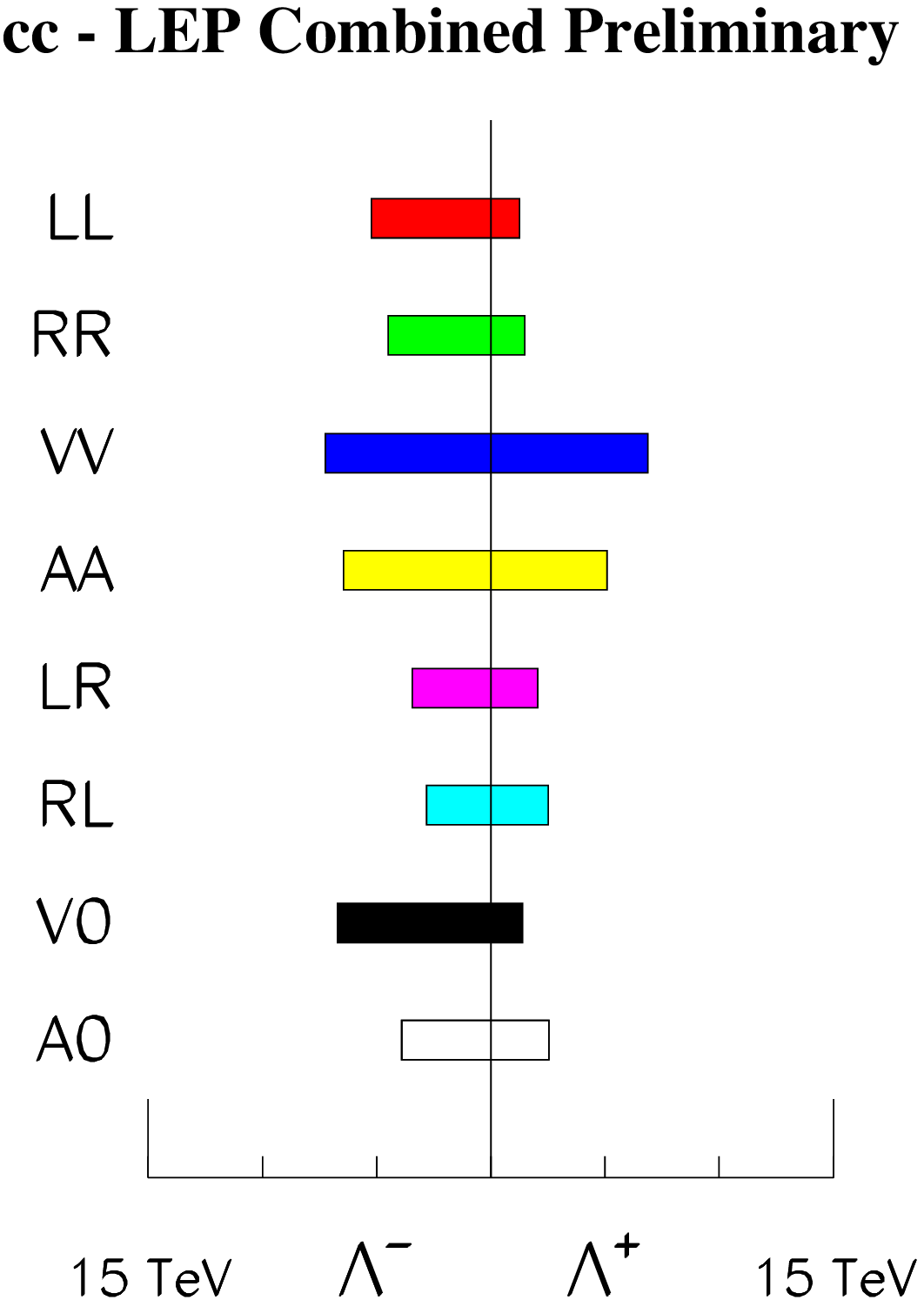,height=6.cm}}
\end{center}
\caption{\label{fig:ci_results} Limits on the energy scale $\Lambda$ for
contact interactions of the form $\rm e^+e^-\rightarrow \ell^+\ell^-$ (left),
$\rm e^+e^-\rightarrow \rm b\bar{\rm b}$ (middle), $\rm e^+e^-\rightarrow
\rm c\bar{\rm c}$ (right) and for different models of contact
interactions interfering destructively ($\Lambda^-$) or
constructively ($\Lambda^+$) with the Standard Model contributions.
These limits are calculated assuming $g^2=4\pi$.
The different bands represent values excluded at 95\% confidence level.}
\end{figure}

The existence of four-fermion contact interactions would result in
deviations  in the di-fermion cross-section and asymmetry measurements
from the Standard Model predictions.  
No deviations are observed.  Limits
on the energy scale $\Lambda$ are calculated for different models by
fitting the measured di-fermion
cross-sections, leptonic forward-backward asymmetries, 
$\rm R_{\rm b}$ and $\rm R_{\rm c}$ at all
the center-of-mass energies achieved at LEP.  
Figure~\ref{fig:ci_results} presents graphically the combined LEP results of
the fit to different models for the contact interactions $\rm
ee\ell\bar{\ell}$, $\rm eeb\bar{b}$ and $\rm
eec\bar{c}$~\cite{bib:LEP_2f_results}.    
These particular interactions
are only accessible at LEP.   
The  limits $\Lambda^-,\Lambda^+$ correspond to destructive and
constructive interference between the Standard Model and contact
interactions contributions.
Limits for $\rm e^+e^-\rightarrow\ell^+\ell^-$ interactions are 
obtained, assuming lepton universality, by combining results from $\rm
e^+e^-\rightarrow \mu^+\mu^-$ and $\rm
e^+e^-\rightarrow \tau^+\tau^-$.

\section{\boldmath{$\rm Z^\prime$} Boson}

Many different models predict the existence of additional gauge
bosons.  Precision measurements of LEP data are used to constrain the
parameters associated with an extra neutral gauge boson 
($\rm Z^\prime$).  In 
general, an additional neutral boson could mix with the Standard Model
$\rm Z^0$ boson.  This mixing can be parameterised  
by the angle $\theta_{\rm ZZ^\prime}$.

Limits on the mass ($\rm M_{\rm Z^\prime}$) and mixing angle ($\theta_{\rm
ZZ^\prime}$) are obtained from measurements, both at the $\rm Z^0$ pole
and at higher energy, of the di-fermion 
cross-sections and leptonic forward-backward asymmetries.
At the $\rm Z^0$ pole, measured cross-sections and asymmetries are
particularly sensitive to the coupling of a possible $\rm
Z^\prime$ to fermions and are therefore most sensitive to the mixing
angle $\theta_{\rm ZZ^\prime}$.
At high energy, the interference between $\rm Z^0$ and $\rm Z^\prime$
is important and the data are consequently particularly sensitive to the 
mass $\rm M_{\rm Z^\prime}$.

\begin{figure}
\begin{center}
\mbox{\epsfig{file=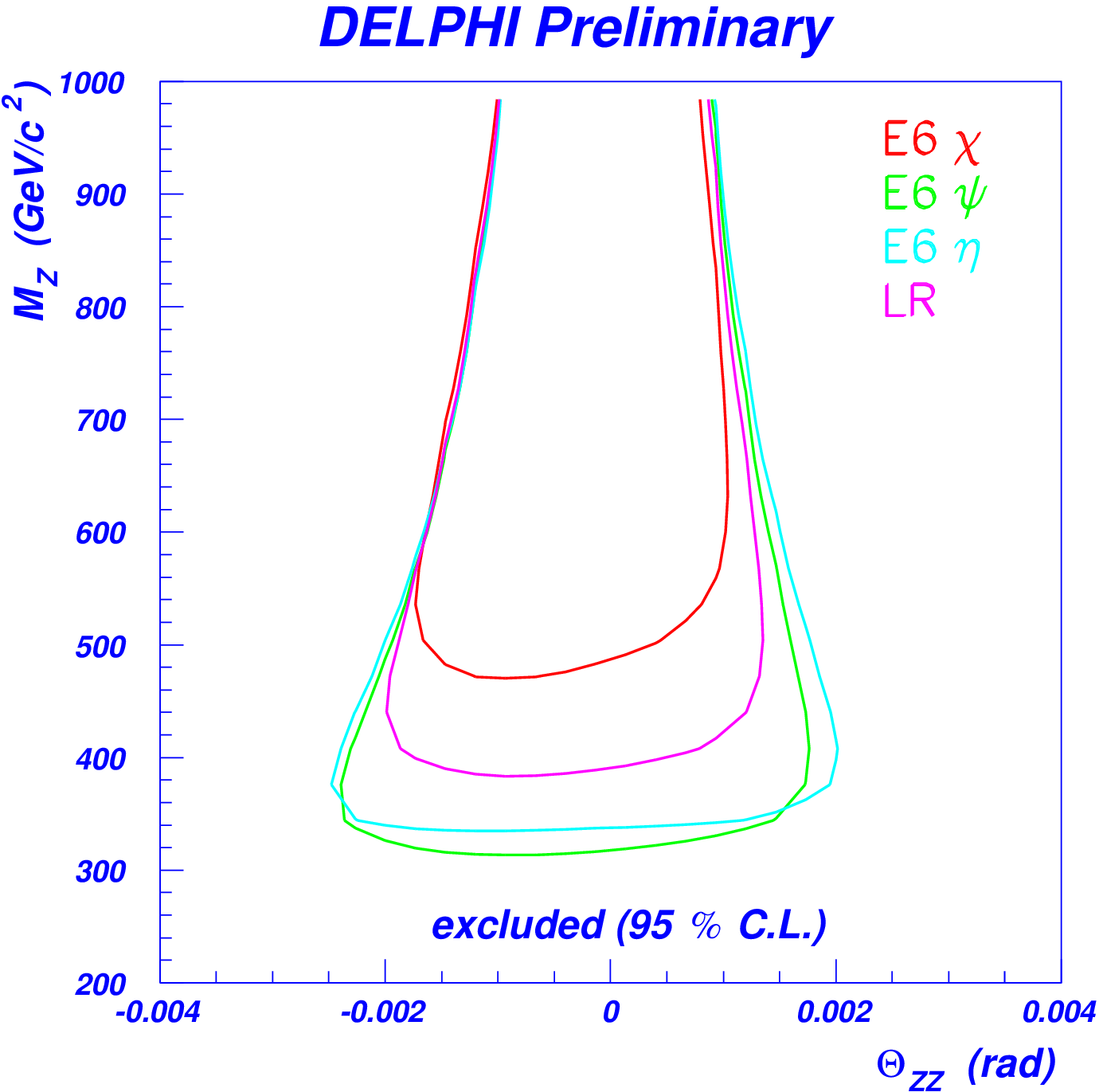,height=8.3cm}}
\hspace{0.5cm}
\mbox{\epsfig{file=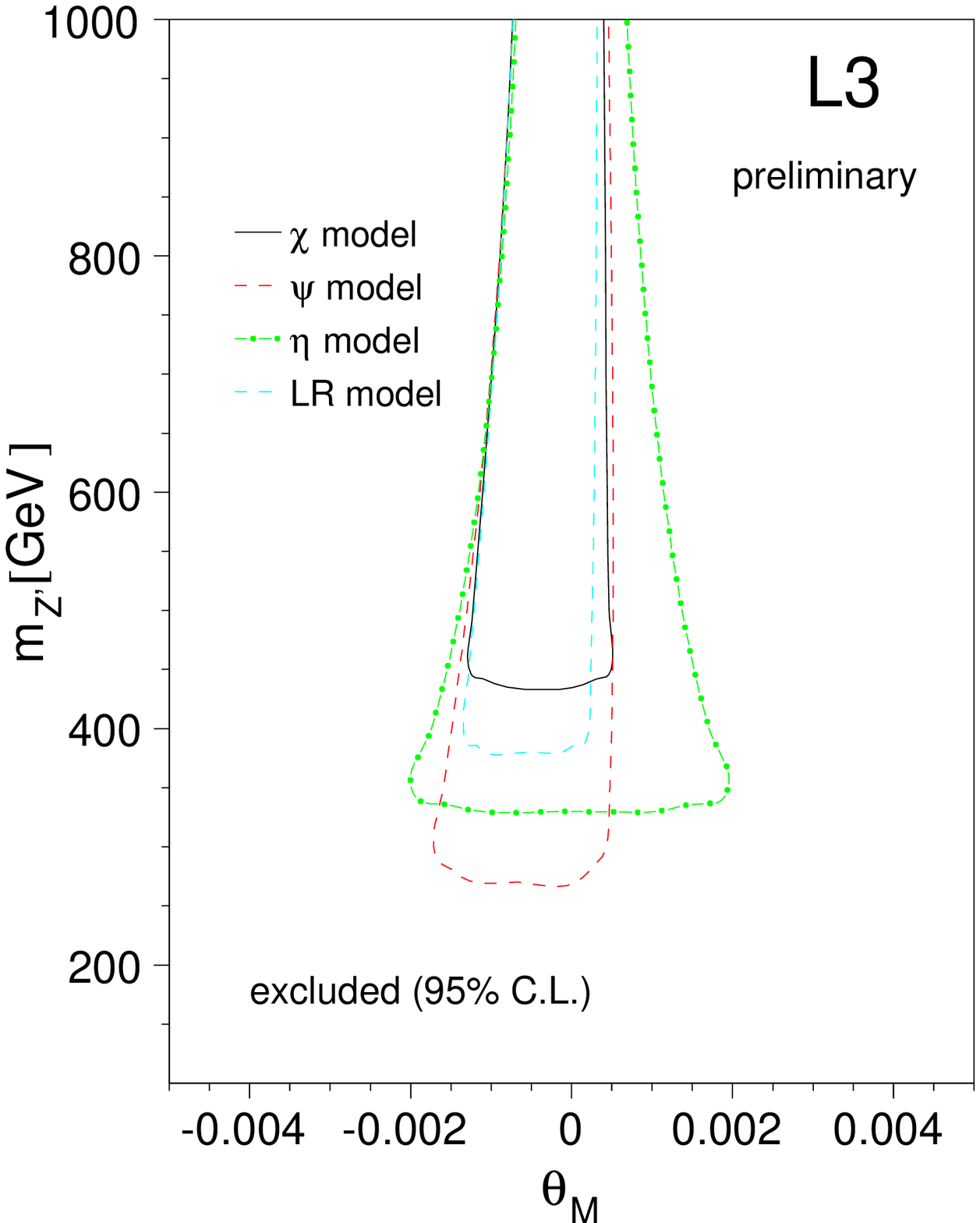,height=8.3cm}}
\caption{\label{fig:zp_limits} Two-dimensional limits on the mass and
mixing angle of $\rm Z^\prime$ boson obtained by DELPHI~\cite{bib:zp_results_DELPHI} and L3~\cite{bib:zp_results_L3} for different models predicting
the existence of an additional neutral gauge boson. These preliminary results
include the entire data set recorded by the respective experiments.
Regions outside the curves are excluded at the 95\% confidence level.}
\end{center}
\end{figure}

Figure~\ref{fig:zp_limits} shows examples of two-dimensional limits
obtained for four different models:  Left-Right
Symmetric~\cite{bib:zp_models,bib:zp_lr}, 
Sequential Standard Model~\cite{bib:zp_ssm} and $\rm E_6$ GUT models ($\chi,\psi,\eta$)~\cite{bib:zp_models,bib:zp_e6}.
Regions outside the curves are excluded at the 95\% confidence level.
Table~\ref{table:zprime} shows the 95\% confidence level lower mass limits
obtained assuming the
$\rm Z^\prime$ is decoupled from the Standard Model $\rm Z^0$ ($\theta_{\rm ZZ^\prime} = 0$).

\begin{table}
\renewcommand{\arraystretch}{1.1}
\renewcommand{\arraycolsep}{2.pt}
\begin{center}
\begin{tabular}{|l|c c r c|r|} \cline{2-6}
\multicolumn{1}{c|}{} & \multicolumn{5}{c|}{95\% CL lower limit on $\rm M_{\rm Z^\prime}$ (GeV)} \\ \cline{2-6}
\multicolumn{1}{c|}{} & ALEPH & DELPHI & \multicolumn{1}{c}{L3} & OPAL & LEP \\ \hline
$\chi$ & $680^*$ & 503 & 460 & 684 & 715 \\ 
$\psi$ & $410^*$ & 336 & 275 & 334 & 478 \\ 
$\eta$ & $350^*$ & 353 & 330 & 451 & 454 \\
LRS & $600^*$ & 412 & 450 & 469 & 862 \\
SSM & --- & --- & 1280 & ---& 2090 \\ \hline
\end{tabular}
\end{center}
\caption{\label{table:zprime} Lower
bounds~\cite{bib:LEP_2f_results,bib:zp_results_DELPHI,bib:zp_results_L3,bib:zp_results_OPAL,bib:zp_results_ALEPH}
on the mass of a $\rm Z^\prime$ boson for different models assuming $\theta_{\rm ZZ^\prime} = 0$.  Results
marked with * only include data up to 202~GeV.  Although not all the
experiments have reported individual mass limits for the SSM model,
the LEP electroweak working group has combined the preliminary
cross-section and asymmetry measurements of each experiment up to
$\sqrt{s}=209$~GeV
and extracted a lower bound on the $\rm Z^\prime$ mass in the context
of the SSM model.  Furthermore, in addition to cross-section and
asymmetry measurements, limits calculated by 
ALEPH include their precision measurement of $\rm R_{\rm b}$. }
\end{table}

\section{Leptoquarks}

Leptoquarks are spin 0 or spin 1 particles carrying both lepton and
baryon numbers and mediating interactions between quarks and leptons.
In general, the total number of possible leptoquark states is reduced to 9
scalar and 9 vector states by
assuming that the interaction between a leptoquark, a quark and a
lepton be dimensionless, $\rm SU(2)\times U(1)$
invariant and conserves baryon/lepton numbers.  Under these
assumptions, the list of possible
leptoquark states can be found in \cite{bib:lq_states}.
Leptoquarks are said to be first generation leptoquarks if they only
couple to first generation leptons and quarks.

\begin{figure}
\begin{center}
\mbox{\epsfig{file=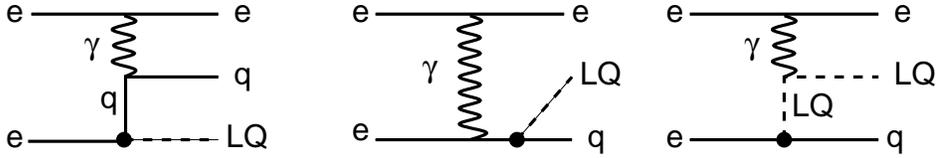,height=2.cm}}
\caption{\label{fig:lq_single_feynman} Dominant diagrams at LEP for
the single production of first generation leptoquarks.}
\end{center}
\end{figure}
\begin{figure}
\begin{center}
\mbox{\epsfig{file=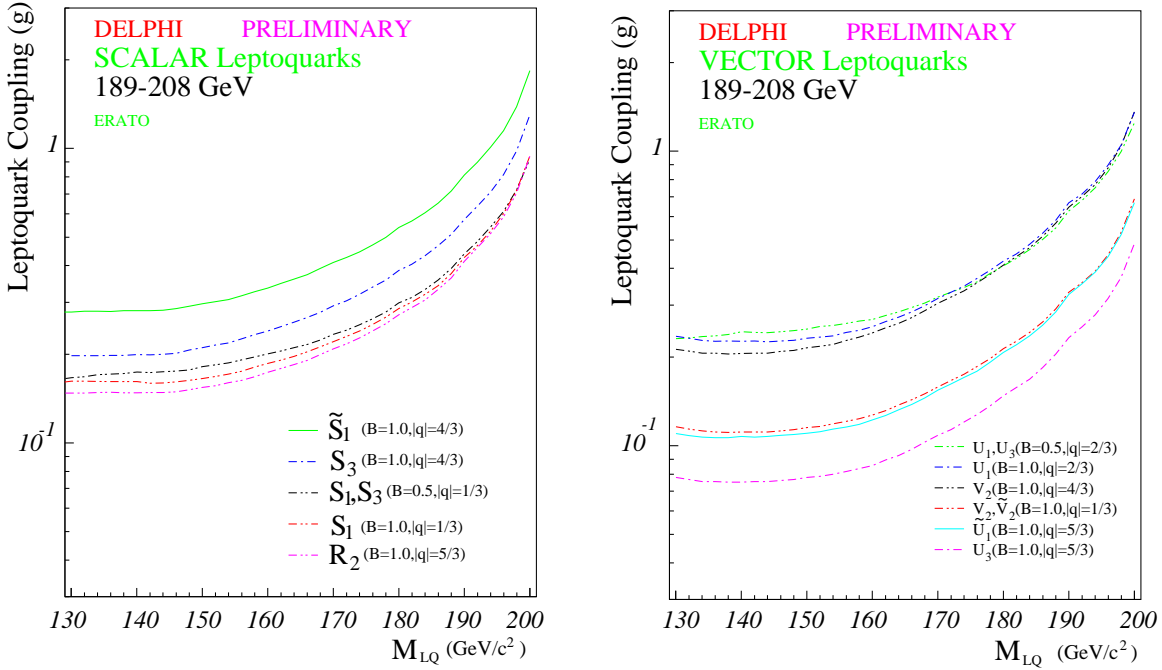,height=9.cm}} 
\caption{\label{fig:lq_single_DELPHI} Limits on leptoquark coupling
parameters as a function of mass obtained by
DELPHI~\cite{bib:lq_single_DELPHI} for different scalar (left) and
vector (right) leptoquark states. 
Regions above the curves are excluded at the 95\% confidence level.}
\end{center}
\end{figure}

First generation leptoquarks could be singly 
produced at LEP 
in association with an electron and a quark which
often travel in the forward direction outside the detector acceptance.
The dominant diagrams for single production of leptoquarks at LEP are
shown in figure~\ref{fig:lq_single_feynman}.
The leptoquarks produced then decay to
a lepton and a quark resulting in event final states containing a monojet, or
one jet and one isolated electron.
The production cross-section depends on the mass of the leptoquark
$\rm M_{\rm LQ}$, and on the coupling parameters $\rm g_{\rm L,R}$, 
where L,R denote the chirality of the lepton it is coupling to.
Data analysed are compatible with expectations from the Standard Model.
Figure~\ref{fig:lq_single_DELPHI} shows an example of limits on the
coupling as a function of leptoquark mass obtained for
different scalar and vector leptoquark states.

The pair production of leptoquarks at LEP would mainly proceed
through an s-channel $\gamma/\rm Z^0$ exchange diagram.  
Given the existing limits on
the couplings  $\rm g_{\rm L,R}$, the t-channel contribution to the
production of first generation leptoquarks is negligible.
Thus, the pair production of leptoquarks is independent of the couplings
$\rm g_{\rm L,R}$ and only depends on the leptoquark masses.
All three generations of leptoquarks could be pair produced.
The decay of both leptoquarks results in event final states containing two
hadronic jets and two leptons, two jets and one lepton, or two jets and
missing energy.
Limits on leptoquark masses obtained
from pair production searches are shown in table~\ref{table:lq_pair}.

\begin{table}
\begin{center}
\renewcommand{\arraystretch}{1.2}
\begin{tabular}[t]{|l r | r r r |}\cline{3-5}
\multicolumn{2}{c}{} & \multicolumn{3}{|c|}{OPAL} \\ 
\multicolumn{2}{c}{} & \multicolumn{3}{|c|}{Mass limits (GeV)}\\ \hline
LQ & $\rm Q_{\rm em}$ & $1^{\rm st}$ & $2^{\rm nd}$ & $3^{\rm rd}$ \\
\hline\hline 
$\rule{0.cm}{0.3cm}\tilde{S}_{\scriptscriptstyle 0}$ & -4/3 & 97.6 & 100.3 & 98.2 \\ \hline
$\rule{0.cm}{0.3cm}S_{\scriptscriptstyle \frac{1}{2}}$ & -2/3 & 90.9 & 90.4 & 87.2 \\ 
 & -5/3 & 98.7 & 100.8 & 99.2 \\ \hline
$\rule{0.cm}{0.3cm}\tilde{S}_{\scriptscriptstyle \frac{1}{2}}$ & +1/3 & 76.5 & 76.5 & 76.5 \\ 
 & -2/3 & 93.9 & 98.6 & 94.9 \\ \hline
$\rule{0.cm}{0.3cm}S_{\scriptscriptstyle 1}$ & +2/3 & 94.7
& 94.7 & 94.7 \\ 
 & -4/3 & 99.2 & 101.1 & 99.6 \\ \hline
\end{tabular}
\hspace{1.7cm}
\begin{tabular}[t]{|l r | r r r |}\cline{3-5}
\multicolumn{2}{c}{} & \multicolumn{3}{|c|}{OPAL} \\ 
\multicolumn{2}{c}{} & \multicolumn{3}{|c|}{Mass limits (GeV)}\\ \hline
LQ & $\rm Q_{\rm em}$ & $1^{\rm st}$ & $2^{\rm nd}$ & $3^{\rm rd}$ \\
\hline\hline 
$\rule{0.cm}{0.3cm}V_{\scriptscriptstyle 0}$ & -2/3 & 98.7 &
 98.7 & 96.5 \\ \hline
$\rule{0.cm}{0.3cm}\tilde{V}_{\scriptscriptstyle 0}$ &
-5/3 & 101.7 & $\scriptscriptstyle >$ 102 & 101.8 \\ \hline
$\rule{0.cm}{0.3cm}V_{\scriptscriptstyle \frac{1}{2}}$ & -1/3 & 98.9 & 98.7 & 98.2 \\ 
 & -4/3 & 101.3 & $\scriptscriptstyle >$ 102 & 101.5 \\ \hline
$\rule{0.cm}{0.3cm}\tilde{V}_{\scriptscriptstyle \frac{1}{2}}$ & +2/3 & 98.3 & 98.3 & 98.3 \\ 
 & -1/3 & 99.9 & 101.4 & 100.3 \\ \hline
 & +1/3 & 100.7 & 100.7 & 100.7 \\ 
$\rule{0.cm}{0.3cm}V_{\scriptscriptstyle 1}$   & -2/3 & 98.7 & 98.7 & 96.5 \\
  & -5/3 & 101.9 & $\scriptscriptstyle >$ 102 & $\scriptscriptstyle >$
102 \\ \hline 
\end{tabular}
\end{center}
\caption{\label{table:lq_pair} Lower mass limits of 
scalar (left) and vector (right)
leptoquarks obtained by OPAL~\cite{bib:searches_OPAL} from the search
of pair produced leptoquarks at $\sqrt{s} = 200-209$~GeV.  
For each leptoquark state considered,
95\% confidence level lower bounds on the mass are given for $1^{\rm
st}, 2^{\rm nd}$ and $3^{\rm rd}$ generation leptoquarks.
The column $\rm Q_{\rm em}$ gives the electric charge
of each leptoquark. }
\end{table}


Leptoquarks could also contribute to the production of two-jet events via a
t-channel leptoquark exchange.
Limits on leptoquark masses in this case are obtained by fitting the
measured $\rm 
e^+e^- \rightarrow q\bar{q}$ cross-section at different centre-of-mass
energies.  The predicted deviations in the
cross-section are calculated assuming one coupling ($\rm g_{\rm L}$ or
$\rm g_{\rm R}$) to be equal to
$\sqrt{4\pi\alpha_{\rm em}}$ and setting the other to zero.
Furthermore, the couplings are assumed to be flavour diagonal.
Table~\ref{table:lq_indirect} presents preliminary mass limits 
for different leptoquark states
obtained from indirect searches. 

\begin{table}
\begin{center}
\renewcommand{\arraystretch}{1.}
\begin{tabular}{|l l|c|} \hline
\multicolumn{2}{|c|}{Scalar LQ} & $\rm M_{\rm LQ}$ \\ 
\multicolumn{2}{|c|}{} & (GeV) \\ \hline
\rule{0.cm}{0.5cm}$\rm S_{\scriptscriptstyle 0}$ & 
$\rm \rightarrow e_{\scriptscriptstyle \rm L}u$ & 789 \\ 
                            & 
$\rm \rightarrow e_{\scriptscriptstyle \rm R}u$  & 639 \\ \hline
\rule{0.cm}{0.5cm}$\rm \tilde{S}_{\scriptscriptstyle 0}$ & 
$\rm \rightarrow e_{\scriptscriptstyle \rm R}u$ & 210 \\ \hline
\rule{0.cm}{0.5cm}$\rm S_{\scriptscriptstyle \frac{1}{2}} $ & 
$\rm \rightarrow e_{\scriptscriptstyle \rm L}\bar{u} $ & 189 \\ 
                                  & 
$\rm \rightarrow e_{\scriptscriptstyle \rm R}\bar{u},e_{\scriptscriptstyle \rm R}\bar{d} $ & 240 \\ \hline
\rule{0.cm}{0.5cm}\raisebox{0.5ex}
{$\rm\tilde{S}_{\scriptscriptstyle \frac{1}{2}}$} & $\rm \rightarrow
e_{\scriptscriptstyle \rm L}\bar{d}$ & - \\ \hline

\rule{0.cm}{0.5cm}$\rm S_{\scriptscriptstyle 1} $ &
 $\rm \rightarrow e_{\scriptscriptstyle \rm L} u,e_{\scriptscriptstyle
 \rm L} d $ & 364 \\ \hline
\end{tabular}
\hspace{1.cm}
\begin{tabular}{|l l|c|} \hline
\multicolumn{2}{|c|}{Vector LQ} & $\rm M_{\rm LQ}$ \\ 
\multicolumn{2}{|c|}{} & (GeV) \\ \hline
\rule{0.cm}{0.5cm}$\rm V_{\scriptscriptstyle 0}$ & 
$\rm \rightarrow e_{\scriptscriptstyle \rm L}\bar{d}$ & 1070 \\
                               &
$\rm \rightarrow e_{\scriptscriptstyle \rm R}\bar{d}$ & 167 \\ \hline
\rule{0.cm}{0.5cm}$\rm \tilde{V}_{\scriptscriptstyle 0}$ & 
$\rm \rightarrow e_{\scriptscriptstyle \rm R}\bar{u}$ & 497 \\  \hline
\rule{0.cm}{0.5cm}$\rm V_{\scriptscriptstyle \frac{1}{2}}$ & 
$\rm \rightarrow e_{\scriptscriptstyle \rm L}d$ & 305 \\
                                 &
$\rm \rightarrow e_{\scriptscriptstyle \rm R}u,e_{\scriptscriptstyle
\rm R}d$  & 227 \\ \hline
\rule{0.cm}{0.5cm}\raisebox{0.5ex}
{$\rm \tilde{V}_{\scriptscriptstyle \frac{1}{2}}$} & 
$\rm \rightarrow e_{\scriptscriptstyle \rm L}u$ & 176 \\  \hline
\rule{0.cm}{0.6cm}$\rm V_{\scriptscriptstyle 1}$ & 
$\rm \rightarrow e_{\scriptscriptstyle \rm L}\bar{u},e_{\scriptscriptstyle \rm L}\bar{d}$ & 664 \\ \hline
\end{tabular}
\caption{\label{table:lq_indirect} Preliminary LEP combined lower
limits~\cite{bib:LEP_2f_results} on leptoquark masses obtained from
fits to the measured 
$\rm e^+e^- \rightarrow q\bar{\rm q}$ cross-section at all
centre-of-mass energies.  Bounds are expressed at the 95\% confidence level.
For the state $\rm\tilde{S}_{\scriptscriptstyle \frac{1}{2}}$, the predicted 
cross-section is too small and no limits can be set. }
\end{center}
\end{table}

\section{Excited Leptons}

Excited states of the known leptons naturally arise in compositeness
 models which assume fermions have substructure.  
At LEP, excited leptons ($\ell^*$) could be both pair or singly produced.
Limits on the mass and coupling of excited leptons are calculated in
the framework of the model described in \cite{bib:ex_models}.
The gauge invariant Lagrangian describing the coupling between an
excited lepton, a lepton and a gauge boson is given by

\[ {\cal L} =\frac{1}{2 {\Lambda}} \bar{\ell}^*
 \sigma^{\mu\nu}\left[g 
 f \frac{ \mbox{\boldmath $\tau$} }{2}
 \mbox{\boldmath $W$}_{\mu\nu} +
 g^\prime f^\prime \frac{Y}{2} B_{\mu\nu} \right] \ell_{\rm
 L} \]

\noindent
where $\sigma^{\mu\nu}$ is the covariant bilinear tensor, \mbox{\boldmath $\tau$}
denotes the Pauli matrices, $Y$ is the weak hypercharge, ${\bf W_{\mu\nu}}$ 
and $B_{\mu\nu}$ represent the Standard Model gauge field tensors and the
couplings $g,g^\prime$ are the SU(2) and U(1)  
coupling constants of the Standard Model.
The compositeness scale is set by the parameter $\Lambda$ which has units 
of energy.  Finally, the strength of the $\ell\ell^*\rm V$ couplings is 
governed by the constants $f$ and $f^\prime$.
These constants can be interpreted as weight factors associated to 
the different gauge groups.
The values of $f$ and $f^\prime$ dictate the relative branching
fractions of excited leptons to each gauge boson. 
In order to reduce the number of free parameters, limits are usually
calculated for the specific cases $f = f^\prime$ and $f=-f^\prime$.


Excited leptons could be pair produced via an s-channel $\gamma/\rm Z^0$
diagram.   Since excited leptons would promptly decay to any
of the gauge bosons, a large number of final states are possible.
Table~\ref{table:ex_pp} summarizes the lower mass
limits obtained from the search of pair produced excited leptons by
all four LEP experiments and for different $f,f^\prime$ assignments.
In general, mass limits from pair production searches are set close to the kinematic limit.  Limits presented in
table~\ref{table:ex_pp} sometimes differ between experiments since
different data sets were used to calculate the limits, see
references~\cite{bib:searches_OPAL,bib:ex_pair_ALEPH,bib:ex_pair_DELPHI,bib:ex_L3,bib:ex_pair_OPAL}
for further details.

\begin{table}
\begin{center}
\renewcommand{\arraystretch}{1.1}
\begin{tabular}{||c||c|c|c|c||c|c||c||} \cline{2-8}
\multicolumn{1}{c||}{} & \multicolumn{7}{c||}{\rule{0cm}{0.4cm}
\normalsize 95\% CL Mass Limit (GeV)} 
                                                               \\ \cline{2-8}
\multicolumn{1}{c||}{} & \multicolumn{4}{c||}{$f=f^\prime$} &
                       \multicolumn{2}{c||}{$f=-f^\prime$} &
                       \multicolumn{1}{c||}{coupling}  \vspace{-0.2cm}\\ 
\multicolumn{1}{c||}{} & \multicolumn{4}{c||}{} &
                       \multicolumn{2}{c||}{} &
                       \multicolumn{1}{c||}{independent} \\ \cline{2-8}
\multicolumn{1}{c||}{} & {DELPHI}     & 
                         {OPAL}   & 
                         {L3} & 
                         {ALEPH}  & 
                         {DELPHI}     & 
                         {L3}   & 
                         {L3}     \\ \hline\hline
{\boldmath$ \rm e^*$} & 103.0         & 
                               102.9         & 
                               100.0 & 
                               94.3 & 
                               98.0 & 
                               93.4          & 
                               93.3          \\ \hline
{\boldmath$ \mu^*$}  &  103.1         & 
                               102.9         &
                               100.2 & 
                               94.3 & 
                               98.0 & 
                               93.4          & 
                               93.4          \\ \hline
{\boldmath$ \tau^*$} &  102.2 & 
                               102.8         & 
                               99.8         & 
                               94.3 & 
                               98.0 & 
                               93.4          & 
                               92.2          \\ \hline\hline
{\boldmath $\nu^*_{\rm e}$} &  102.0 & 
                               99.5          & 
                               99.1          & 
                               94.2 & 
                               102.7 & 
                               99.4          & 
                               98.2          \\ \hline
{\boldmath $\nu_\mu^*$} &  102.4 & 
                               99.5          & 
                               99.3          &                 
                               94.2 & 
                               102.8 & 
                               99.4          & 
                               98.3          \\ \hline
{\boldmath $\nu_\tau^*$} &  95.3 & 
                               91.9          & 
                               90.5          &   
                               94.2 & 
                               102.8 & 
                               99.4          & 
                               87.8          \\ \hline
\end{tabular}
\end{center}
\caption{\label{table:ex_pp} Lower mass limits obtained from the
searches of pair produced excited leptons and 
for different assignments of $f$ and $f^\prime$.  
Bounds from ALEPH~\cite{bib:ex_pair_ALEPH}, DELPHI~\cite{bib:ex_pair_DELPHI}, L3~\cite{bib:ex_L3} and
OPAL~\cite{bib:searches_OPAL,bib:ex_pair_OPAL} are calculated using different data sets.}
\end{table}

\begin{figure}
\begin{center}
\raisebox{1.8ex}{\epsfig{file=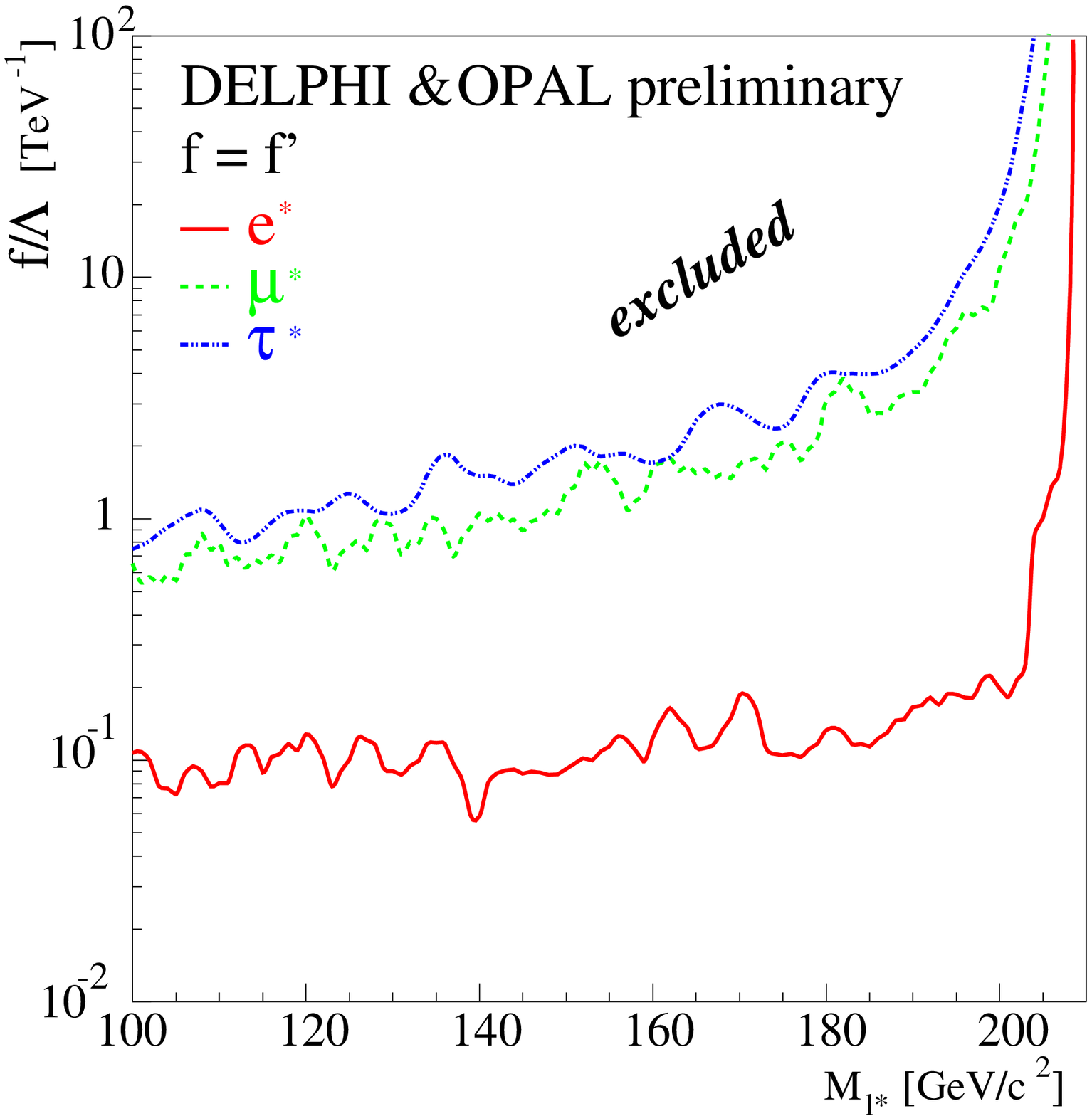,width=4.8cm}}
\mbox{\epsfig{file=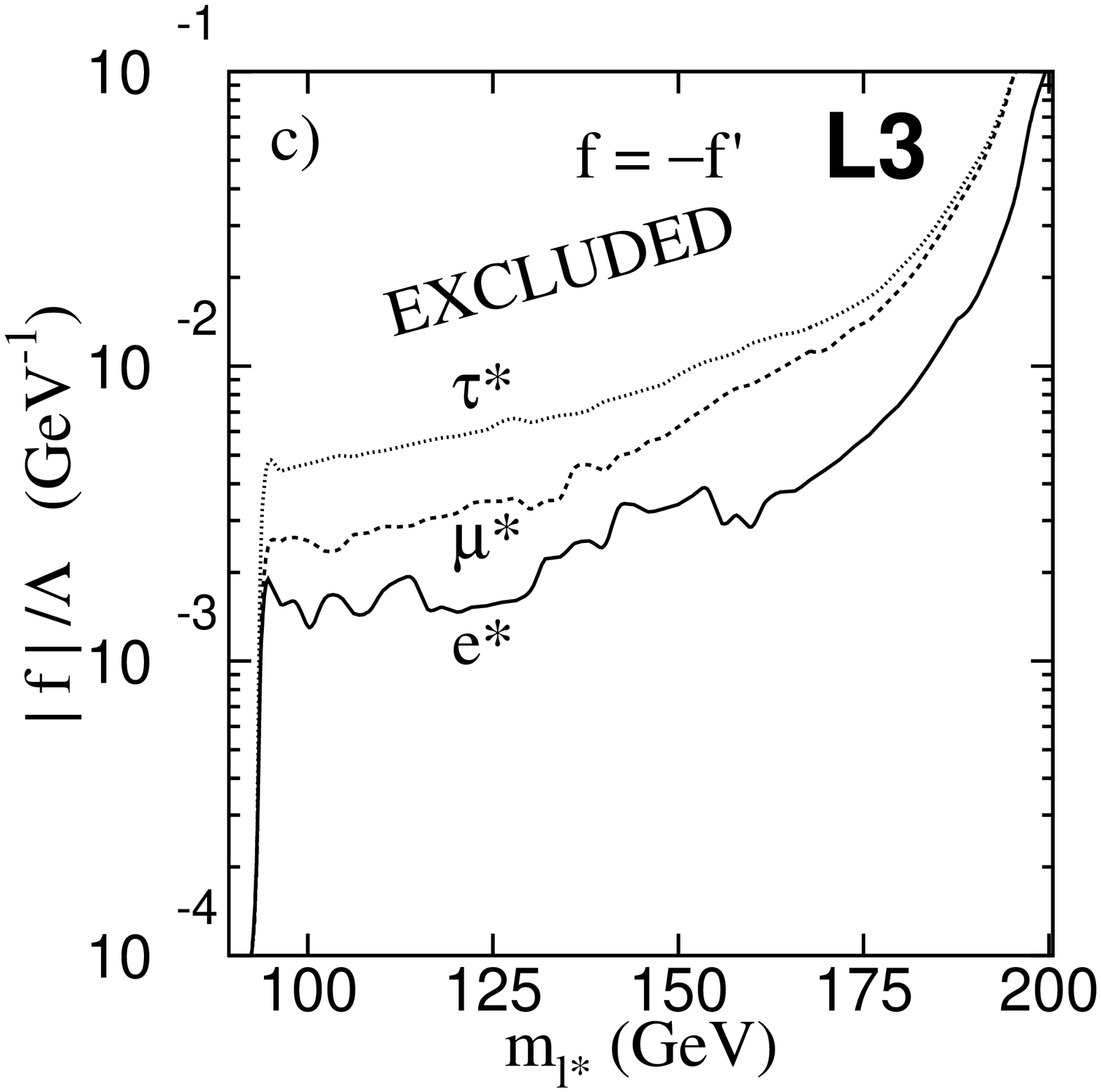,width=5.4cm}}
\mbox{\epsfig{file=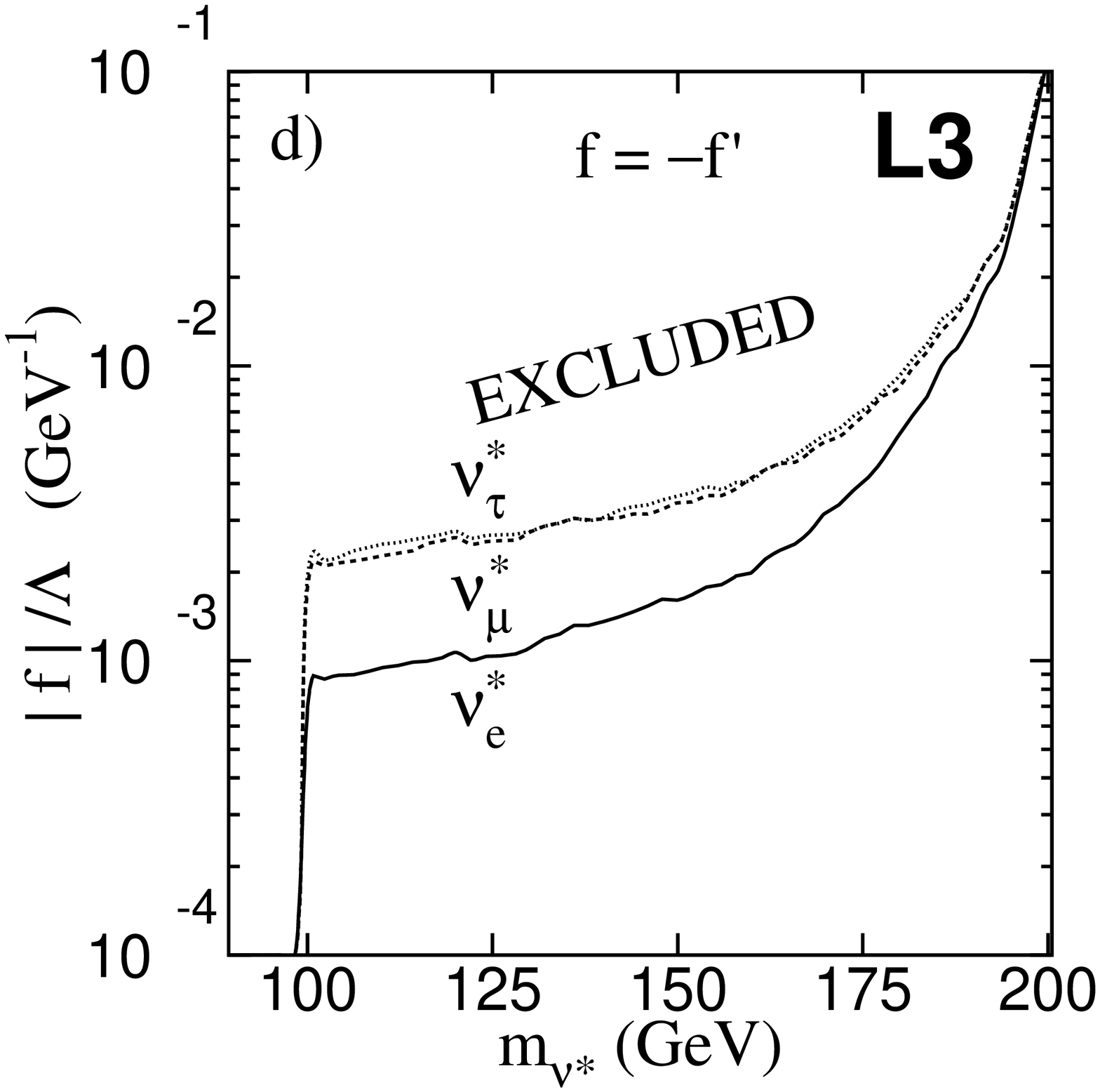,width=5.4cm}}
\end{center}
\caption{\label{fig:ex_limits} Limits on the ratio of the coupling to
the compositeness scale ($f/\Lambda$) as a function of excited lepton
mass for different $f$ and $f^\prime$ assignments.  Regions above the
curves are excluded at the 95\% confidence level.  The combined results~\cite{bib:ex_single_LEP}
from DELPHI and OPAL (left) include data recorded at
$\sqrt{s}=189-208$~GeV while limits from L3~\cite{bib:ex_L3}
(center, right) were 
calculated using data at $\sqrt{s}=192-202$~GeV and are expressed in
$\rm GeV^{-1}$.   }
\end{figure}

Excited leptons could also be singly produced in association with a
Standard Model lepton.  
Possible event final states may contain isolated
leptons and one photon, pairs of hadronic jets and in some
cases missing energy from neutrinos.
No excesses of events are observed in the data.
Limits on the ratio of the coupling to the compositeness scale
($f/\Lambda$) as a function of excited lepton mass are calculated.
Figure~\ref{fig:ex_limits} shows
typical limits obtained from the single production searches for
different $f$ and $f^\prime$ assignments.  
The predicted single production cross-section of excited electrons is enhanced
by the contribution of a t-channel exchange diagram resulting in more
stringent limits than for excited muons and taus.

\section{Technicolour}
Technicolour is a new strong QCD-like interaction that could naturally
induce electroweak symmetry breaking.
In this framework, the $W^\pm$ and $\rm Z^0$ bosons are condensates of a new family
of fermions called technifermions.
The phenomenology associated to the existence of technicolour is
inferred using the framework of the technicolour ``straw-man'' model~\cite{bib:tc_smm}, a
low energy approximation of technicolour containing a minimum number of
free parameters.

Searches for the production of technirho ($\rho_{\rm T}$) and technipions
($\pi_{\rm T}$) are
performed.  The dominant processes searched for include
\[
\begin{array}{l l}
\rm e^+e^- \rightarrow \rho_{\rm T}^{(*)} \rightarrow & \pi^+_{\rm
T}\pi^-_{\rm T} \\
 & \pi^\pm_{\rm T}W^\mp_{\rm L} \\
 & W^\pm_{\rm L}W^\mp_{\rm L} \\
 & \pi^0_{\rm T}\gamma \\
 & \rm f\bar{\rm f} \\
\end{array}
\]
where $W_{\rm L}$ indicates the longitudinal component of the $W$ boson.
The cross-section of the three first final states shown above only
depend on $\rm M_{\rho_{\rm T}}$, $\rm M_{\pi_{\rm T}}$ and $\rm
N_{\rm D}$, the number of technicolour doublets.  For $\rm
M_{\rho_{\rm T}} > 2 M_{\pi_{\rm T}}$, the $\pi^+_{\rm T}\pi^-_{\rm
T}$ final state dominates.  In addition to the technirho exchange diagram, the
reaction $\rm e^+e^- \rightarrow \pi^0_{\rm T}\gamma$ can also proceed
via the exchange of a techniomega ($\omega_{\rm T}$) and is therefore model
dependent.  The cross-section $\rm e^+e^- \rightarrow \pi^0_{\rm
T}\gamma$ depends, in addition to $\rm M_{\rho_{\rm T}}$, $\rm
M_{\pi_{\rm T}}$ and $\rm N_{\rm D}$, on the mass scale $\rm M_{\rm V}$, the
technicolour coupling constant $\alpha_{\rho_{\rm T}}$ and the sum of
the charges of the U and D techniquarks ($\rm Q_{\rm U}+Q_{\rm D}$).
Finally, the $\rm f\bar{\rm f}$ final state is in general suppressed.
Many of the decay final states considered lead to topologies
similar to the Standard Model production of $W$ pairs. 
Also, since approximately
90\% of technipions decay to b quarks, b-tagging is an important part
of the search for technicolour.  

\begin{figure}
\begin{center}
\mbox{\epsfig{file=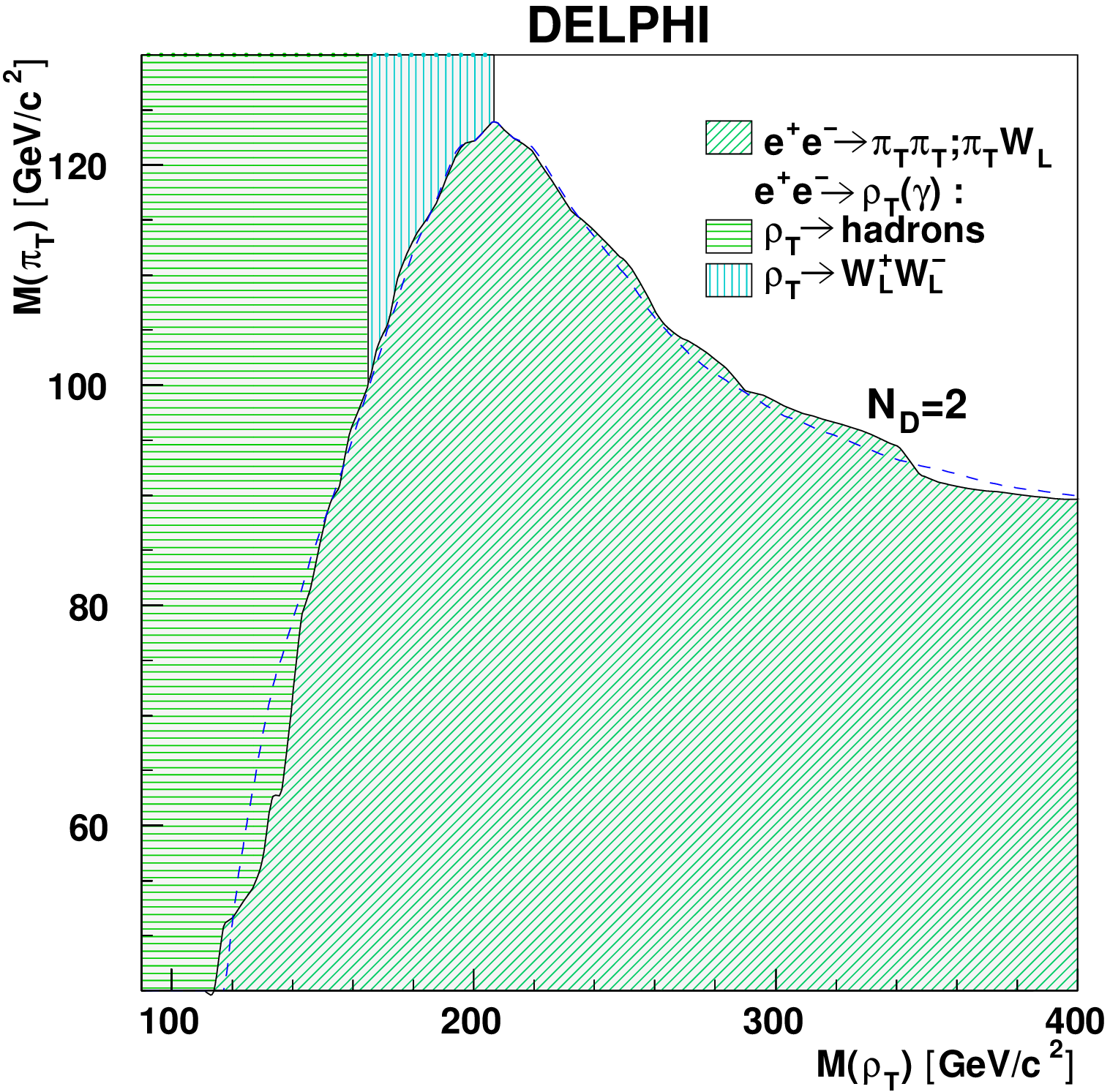,height=7.cm}} \hspace{0.2cm}
\raisebox{-2.6ex}{\epsfig{file=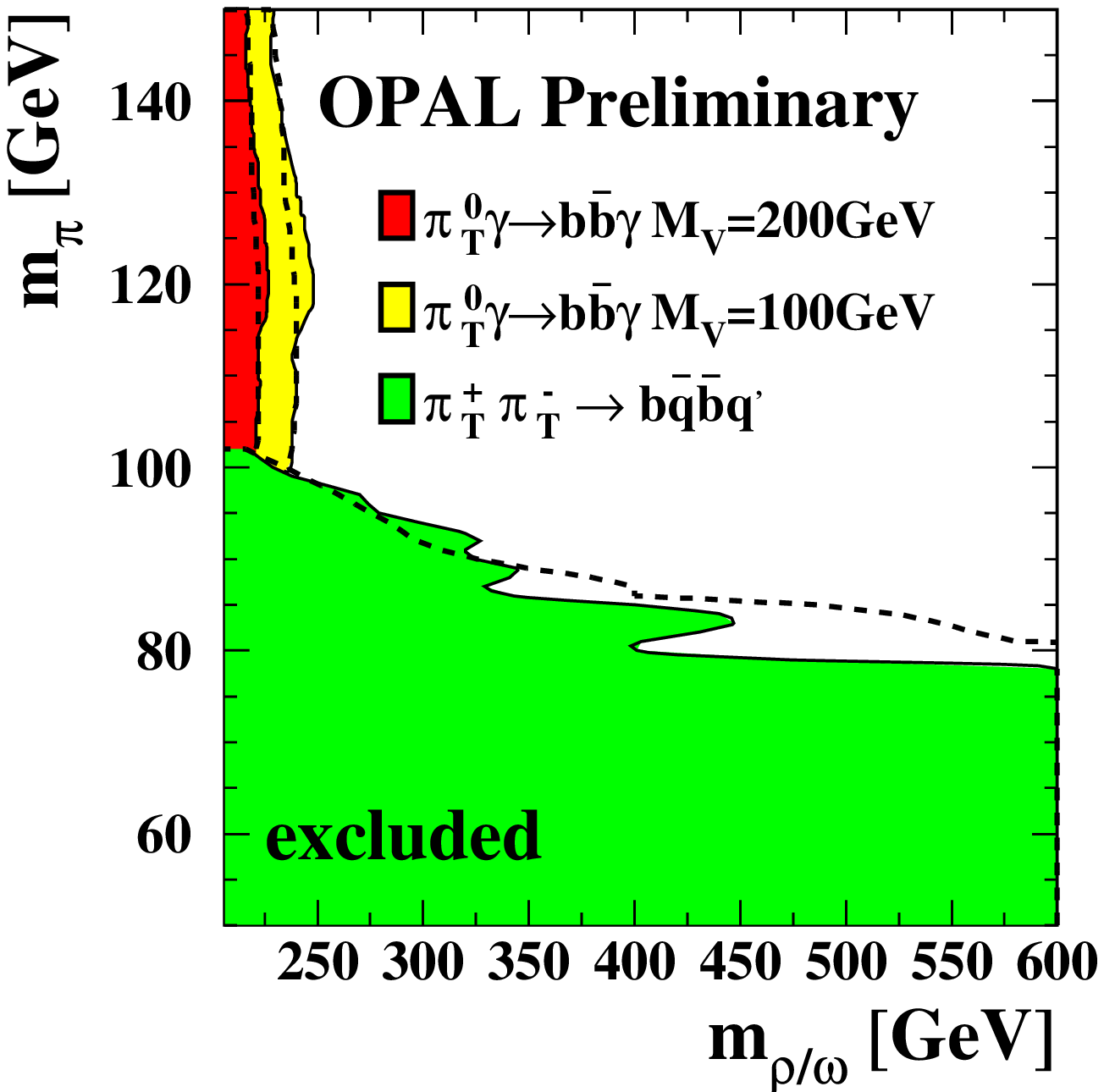,height=8.3cm}}
\vspace{0.9cm} \\
\mbox{\epsfig{file=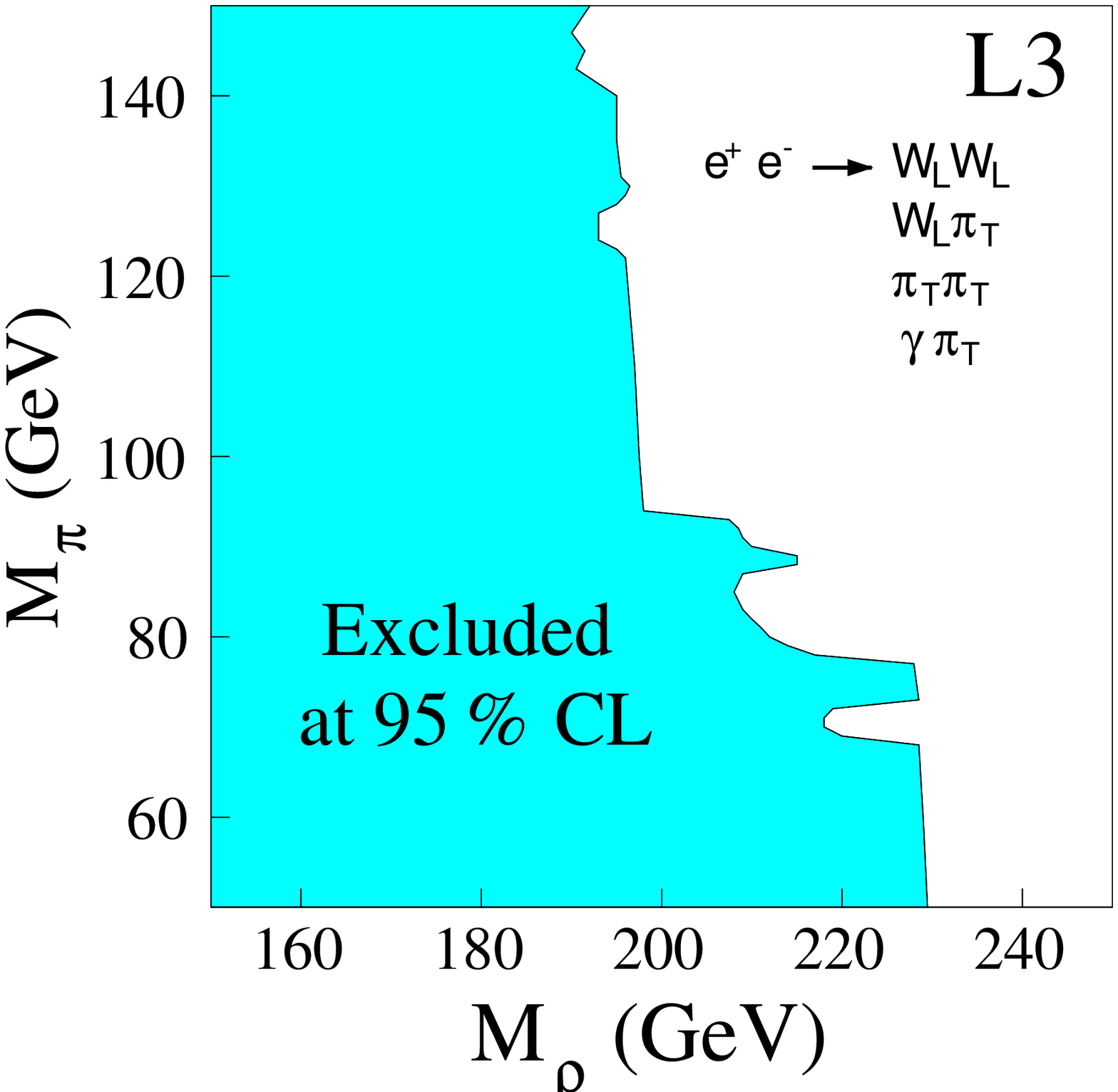,height=7.cm}}\rule{1.5cm}{0.cm}
\end{center}
\caption{\label{fig:tc} Excluded regions in the ($\rm M_{\pi_{\rm T}}$,$\rm
M_{\rho_{\rm T}}$) plane obtained by DELPHI
($\sqrt{s}=192-208$~GeV, top left), OPAL
($\sqrt{s}=200-209$~GeV, top right) and 
L3 ($\sqrt{s}=189$~GeV, bottom). Grey areas are excluded at
95\% confidence level. }
\end{figure}

In the final states considered, no hint of the existence of
technicolour is found. 
Examples of exclusion regions in the ($\rm M_{\pi_{\rm T}}$, $\rm
M_{\rho_{\rm T}}$) plane obtained at LEP are shown in figure~\ref{fig:tc}.
The grey regions are excluded at the 95\% confidence level.
Limits presented from DELPHI~\cite{bib:tc_DELPHI} (top left) were
calculated for the maximal $W_{\rm L}-\pi_{\rm T}$ mixing case,
$\rm N_{\rm D} = 2$.  Figure~\ref{fig:tc} (top right) shows the exclusion
regions obtained by OPAL~\cite{bib:tc_OPAL} for the theoretically
preferred value $\rm N_{\rm D} = 9$, and assuming $(\rm Q_{\rm
U}+Q_{\rm D}) = 5/3$. 
Finally, bounds presented from L3~\cite{bib:tc_L3} (bottom) are valid
for any value of 
$\rm N_{\rm D}$ and for $(\rm Q_{\rm U}+Q_{\rm D}) = 0,1,5/3$.  Limits
on the technipion mass can be extracted from these exclusion plots and
are summarised in table~\ref{table:tc_results}.

\begin{table}
\begin{center}
\renewcommand{\arraystretch}{1.1}
\vspace{1.cm}
\begin{tabular}{|c|c|c|} \cline{2-3}
\multicolumn{1}{c}{} &\multicolumn{2}{|c|}{ 95\% CL limits on $\rm
M_{\pi_{\rm T}}$ (GeV)} \\ \hline
$\rm N_{\rm D}$ & DELPHI & OPAL \\ \hline\hline
2 	& 79.8 	& 62.0 \\ \hline
9 	& 89.1  & 77.0  \\ \hline
\end{tabular}
\end{center}
\caption{\label{table:tc_results} Lower limits on $\rm M_{\pi_{\rm
T}}$ for different values of $\rm N_{\rm D}$ obtained by DELPHI~\cite{bib:tc_DELPHI} and OPAL~\cite{bib:tc_OPAL}. }
\end{table}

\section{Gravity in Extra Dimensions}

Recently, a new scenario has been proposed which tries to 
explain the large difference between the energy scale of gravity and
the other forces. 
The framework developped by Arkani-Hamed, Dimopoulos and
Dvali~\cite{bib:lsg_add} assumes that the universe
is made of extra dimensions in which Standard Model particles are confined to
a 4-dimensional ``wall''.  In this scenario, only gravitons can
propagate in the extra dimensions.  
Furthermore, the Planck mass in 4
dimensions ($\rm M_{\rm Pl}$) can be related to the Planck mass in 4+n
dimensions ($\rm M_{\rm D}$) by the relation $\rm M_{\rm Pl}^2 =
R^{\rm n} M_{\rm D}^{\rm n+2}$, where n is the number of extra
dimensions and R is the size of these extra dimensions.  
The hierarchy between the scale of gravity and the other forces is
removed under the assumption that $\rm M_{\rm D}$ is similar to the
electroweak scale.  
Under these circumstances, extra dimensions could
be ``large'' enough to be probed in collider experiments.
The existence of low scale gravity in extra dimensions would manifest
itself by the direct and indirect production of gravitons.
In 4-dimensions, gravitons would appear as massive, non-interacting
stable particles with a wide range of possible masses.

Gravitons could be produced in the reaction $\rm e^+e^-\rightarrow
G\gamma$ which would result in single photon final states with
missing energy.  The cross-section for such process is proportional to
$(\sqrt{s}/\rm M_{\rm D}^{\rm 2})^{\rm n}$.
Data are in good agreement with the Standard Model
predictions and thus limits on the value of the Planck mass in 4+n
dimensions, $\rm M_{\rm D}$, are calculated.
Table~\ref{table:lsg_direct} shows the lower limits on $\rm M_{\rm D}$
obtained by each LEP experiment as a function of the number of extra
dimensions.


\begin{table}
\renewcommand{\arraystretch}{1.}
\begin{center}
\begin{tabular}{|l | c c c c c c |} \cline{2-7}
\multicolumn{1}{c}{} & \multicolumn{6}{|c|}{Limits on $\rm M_{\rm D}$ (TeV)\rule{0.cm}{0.4cm}} \\ \cline{2-7}
\multicolumn{1}{c}{} & \multicolumn{6}{|c|}{Number of extra dimensions} \\
\multicolumn{1}{c|}{}    & 2   & 3     & 4     & 5     & 6     & 7 \\ \hline
{\bf ALEPH} (189-209 GeV)      & 1.28 & 0.97 & 0.78 & 0.66 & 0.57  & ---\\
{\bf DELPHI} (181-209 GeV)     & 1.38 & --- & 0.84 & --- & 0.58  & ---  \\
{\bf L3} (189 GeV)      & 1.02  & 0.81  & 0.67  & 0.58  & 0.51  & 0.45  \\
{\bf OPAL} (189 GeV)    & 1.09 & 0.86 & 0.71 & 0.61 & 0.53 & 0.47 \\ \hline
\end{tabular}
\end{center}
\caption{\label{table:lsg_direct} Lower bounds on $\rm M_{\rm D}$ from
ALEPH~\cite{bib:lsg_direct_ALEPH},
DELPHI~\cite{bib:lsg_direct_DELPHI}, L3~\cite{bib:lsg_direct_L3} and
OPAL~\cite{bib:lsg_direct_OPAL}.  Limits are calculated at the 95\%
confidence level.}
\end{table}

\begin{figure}
\begin{center}
\mbox{\epsfig{file=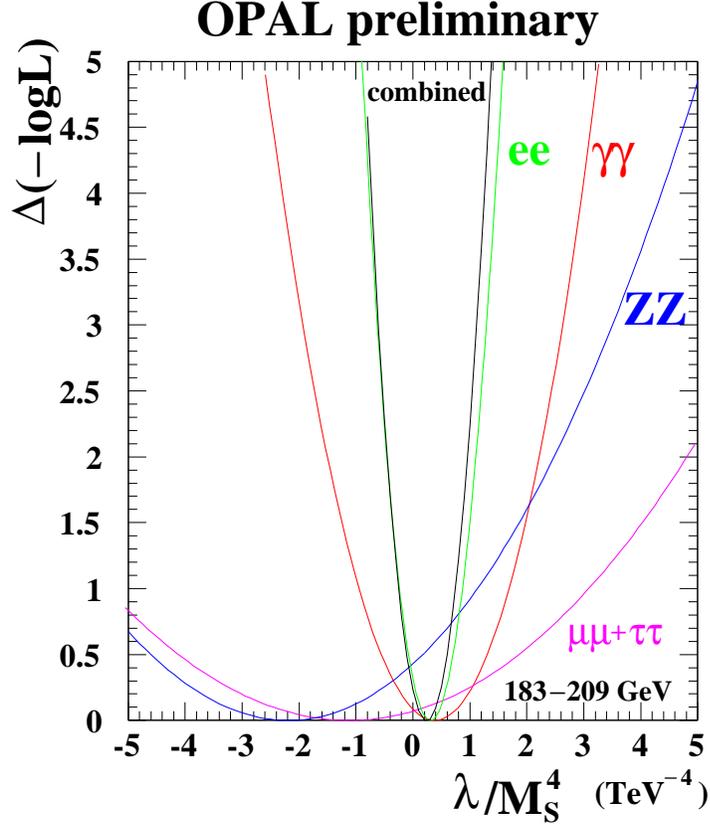,height=11.0cm}}
\end{center}
\caption{\label{fig:lsg_likelihood} Likelihood curves obtained from a
fit to OPAL data recorded at
$\sqrt{s}=183-209$~GeV~\cite{bib:lsg_indirect1_OPAL} for different
final states studied.   }
\end{figure}

Deviations due to the existence of low scale gravity could also be
observed in the di-fermion and di-boson production cross-sections.
The graviton contribution ($\rm e^+e^-\rightarrow G^* \rightarrow
f\bar{\rm f},VV$)  can be parameterised as a function of $\rm
\lambda/M_{\rm s}^4$ where $\lambda$ depends on the details of the
theory of quantum gravity (number of extra dimensions,
compactification, etc.) and $\rm M_{\rm s}$ is a UV cut-off set to the
string scale, presumed to be $\cal
O$(TeV) in these models~\cite{bib:lsg_indirect_cutoff}.  
The di-electron final
state is the most sensitive 
channel to possible contributions from graviton exchange.
Figure~\ref{fig:lsg_likelihood} shows an example of likelihood curves
for different final states studied, 
obtained from a fit to data recorded at $\sqrt{s}=183-209$~GeV
by OPAL .
Resulting limits on $\rm M_{\rm s}$ obtained by
some of the LEP experiments can be found in
table~\ref{table:lsg_indirect}
for $\rm \lambda/M_{\rm s} = \pm 1$.

\begin{table}
\begin{center}
\renewcommand{\arraystretch}{1.1}
\begin{tabular}[b]{| c| c c | }
\multicolumn{1}{c}{}  & \multicolumn{2}{c}{\vspace{0.3cm}\large $e^+e^- \rightarrow e^+e^-$} \\ \cline{2-3}
\multicolumn{1}{c|}{}  & \multicolumn{2}{c|}{Limits on $\rm M_s$ (TeV)}\\
\multicolumn{1}{c|}{}  & $\lambda = -1$ & $\lambda = +1$ \\ \hline
L3   &  0.98 & 1.06 \\ 
OPAL &  1.00 & 1.15 \\ \hline
\end{tabular}
\hspace{1.5cm}
\begin{tabular}[b]{|c |c c |}
\multicolumn{1}{c}{}  & \multicolumn{2}{c}{\vspace{0.3cm}\large $e^+e^- \rightarrow \gamma\gamma$} \\  \cline{2-3}
\multicolumn{1}{c|}{}  & \multicolumn{2}{c|}{Limits on $\rm M_s$ (TeV)}\\
\multicolumn{1}{c|}{}  & $\lambda = -1$ & $\lambda = +1$ \\ \hline
DELPHI & 0.70 & 0.77 \\ 
L3   &  0.99 & 0.84 \\ 
OPAL &  0.89 & 0.83 \\ \hline
\end{tabular}
\end{center}
\caption{\label{table:lsg_indirect} Lower bounds on $\rm M_{\rm s}$
assuming $\rm \lambda = \pm 1$ obtained by DELPHI~\cite{bib:lsg_indirect_DELPHI}, L3~\cite{bib:zp_results_L3,bib:lsg_indirect_L3} and
OPAL~\cite{bib:lsg_indirect1_OPAL,bib:zp_results_OPAL}.  The 95\%
confidence level limits are calculated using $\rm e^+e^- \rightarrow
e^+e^-$ (left) and  $\rm e^+e^- \rightarrow \gamma\gamma$ (right)
event final states.} 
\end{table}

\section{Summary}
Recent preliminary results on various searches performed at LEP have been
summarised.  These represent only a subset of the work involved in the
pursuit of physics beyond the Standard Model.
Much work remains to finalise preliminary results from individual
experiments and publish the final LEP combined limits of many searches.
Data analysed by the four LEP experiments
are in agreement with the Standard Model predictions.


\end{document}